\documentclass[Journal]{IEEEtran}

\usepackage{cite}
\usepackage{colortbl,booktabs}
 \usepackage{xcolor}
 \usepackage{multirow}
 \usepackage{graphicx}
 \usepackage{lscape}
\usepackage{enumitem}
\usepackage{subfig}
\usepackage{amsmath, amsthm, amssymb, amsfonts, mathtools}
\usepackage{bm}
\usepackage{textcomp}
\usepackage{xcolor}
\usepackage{color}
\usepackage{todonotes}
\usepackage{footnote}
\makesavenoteenv{tabular}
\usepackage{tikz}

\makesavenoteenv{table}
\makeatletter
\newcommand\numberthis{\addtocounter{equation}{1}\tag{\theequation}}

\title{DT2CAM: A \underline{D}ecision \underline{T}ree to \underline{C}ontent \underline{A}ddressable \underline{M}emory Framework}

\author{Mariam Rakka, Mohammed E. Fouda, Rouwaida Kanj, and Fadi Kurdahi 
\thanks{M. Rakka and F. Kurdahi are with Center for Embedded \& Cyber-physical Systems, University of California-Irvine, Irvine, CA, USA 92697-2625}
\thanks{M. Fouda is with Center for Embedded \& Cyber-physical Systems, University of California-Irvine, Irvine, CA, USA 92697-2625 and  is also with Nanoelectronics Integrated Systems Center (NISC), Nile University, Giza, Egypt.}
\thanks{R. Kanj is with the ECE Dept., American University of Beirut, Lebanon, 1107 2020} 
\thanks{Manuscript received xxxx xx, xxxx; revised xxxx xx, xxxx.}

}

\begin{document}


\maketitle

\begin{abstract}
Decision trees are considered one of the most powerful tools for data classification. Accelerating the decision tree search is crucial for on-the-edge applications that have limited power and latency budget.   
In this paper, we propose a Content Addressable Memory (CAM) Compiler for Decision Tree (DT) inference acceleration. We propose a novel "adaptive-precision" scheme that results in a compact implementation and enables an efficient bijective mapping to Ternary Content Addressable Memories while maintaining high inference accuracies. In addition, a Resistive-CAM (ReCAM) functional synthesizer is developed for mapping the decision tree to the ReCAM and performing functional simulations for energy, latency, and accuracy evaluations.
We study the decision tree accuracy under hardware non-idealities including device defects, manufacturing variability, and input encoding noise. We test our framework on various DT datasets including \textit{Give Me Some Credit}, \textit{Titanic}, and \textit{COVID-19}.
Our results reveal up to {42.4\%} energy savings and up to 17.8x  better energy-delay-area product compared to the state-of-art hardware accelerators, and up to 333 million decisions per sec for the pipelined implementation. 
\end{abstract}


\section{Introduction}
Machine Learning (ML) continues to play a crucial role in performing complex tasks that are characterized by "learnable" properties. While brain-inspired Deep Neural Networks (DNN) are nowadays thriving in several fields including computer vision, autonomous driving, the Internet of Things (IoT), and smart industries, they are not applicable where interpretability and domain knowledge are required \cite{lecun2015deep,seifert2017visualizations,hernavs2018deep,bengio2007scaling,gubbi2013internet,arrieta2020explainable}. Some applications that require integrating hand-crafted solutions (and hence domain expertise and explainability) as part of the learning process include predictive maintenance, risk management, anomaly detection and image recognition for purposes of medical diagnosis \cite{susto2014machine, bussmann2021explainable,song2018exad,magesh2020explainable}. In particular, Decision Trees (DTs) are popular to perform explainable ML \cite{franklin2005elements,lundberg2020local}, this is known as DT-based ML.

Several hardware accelerators for DT-based ML are proposed in  \cite{saqib2013pipelined,zoulkatni2020hardware,pedretti2021tree,struharik2015decision,chen2011visual,van2012accelerating}.
Most of these are CPU (e.g., Intel X5560), GPU (e.g., Nvidia Tesla M2050), FPGA (e.g., Xilinx Virtex-6), or ASIC-based accelerators. More recently, hardware accelerators based on emerging memories like In-Memory Computing (IMC) architectures have been proposed for DT-based ML \cite{pedretti2021tree, yin2021deep,kang201819}. IMC architectures are gaining momentum for ML applications for they eliminate the memory wall, a known drawback in von Neumann architectures \cite{foster1976content}. Ternary Content Addressable Memories (TCAMs), which perform massively parallel search operations, are considered one realization of IMC architectures
that have proven to boost performance in terms of energy and latency \cite{guo2011resistive,graves2020memory,guo2013ac,arsovski2003ternary, fouda2022memory}.

DT graphs consist of paths (i.e., routes) that describe some rules on features and that terminate by leaf nodes storing class values. To perform inference on decision trees, the incoming data should "match" one single complete path to associate it with some output class. Classical architectures will perform sequential searches on the DT routes to find the matching one. Motivated by the fact that each route in a DT can be mapped to a TCAM row (where the route's feature rules are stored) and by the high search throughput offered by TCAMs, we propose DT2CAM: a Decision Tree to Content Addressable Memory framework. DT2CAM simulates the inference of decision trees on CAMs in general and resistive-based TCAMs in particular. Our {contributions} can be summarized as follows.

\begin{enumerate}
  \item We propose DT2CAM, a framework that bijectively maps any decision tree into TCAM units relying on a novel adaptive precision encoding scheme. DT2CAM comprises two components: 
  \begin{enumerate}
      \item A decision tree to CAM-based hardware architecture compiler (DT-HW compiler). The DT-HW compiler translates a decision tree graph to a structured Look-Up Table (LUT) comprising $0$, $1$, and "don't care" bits. The LUT rows represent encoded DT paths, and they can be mapped into any Ternary CAM architecture.
      \item A ReCAM functional synthesizer which maps the encoded LUT entries to resistive TCAM cells while taking into consideration design requirements and specifications. It also performs simulations to evaluate energy, latency, and accuracy.
  \end{enumerate}
  \item We study the robustness of the proposed DT2CAM framework given hardware non-idealities: manufacturing variability, device defects, and noise in the input dataset. Results prove high robustness characterized by a low accuracy drop compared to recognition accuracy.
  \item We demonstrate for our proposed framework up to {42.4\%} reduction in energy dissipation compared to the similar state-of-the-art hardware accelerator on analog CAMs \cite{pedretti2021tree}, high throughput, and low area overhead. Moreover, we defined a figure of merit (FOM) that further shows that our proposed framework performs the best compared to the other frameworks. 
 We also validate that the DT2CAM functional accuracy matches that of Python-based DT inference.
\end{enumerate}

The rest of the paper is organized as follows. In section II, we explain the proposed DT2CAM framework. Section III presents the implementation details, and Section IV elaborates on the results and compares the framework against other hardware accelerators. Finally, section V concludes the work.


\section{Proposed DT2CAM Framework}
Our proposed framework comprises two components: DT-HW compiler and ReCAM functional synthesizer. The DT-HW compiler translates a decision tree graph to a structured lookup table. The ReCAM functional synthesizer first maps the look-up table into ReCAM arrays and then evaluates energy, latency, and accuracy via simulations.

\subsection{DT-HW Compiler}
To map a decision tree graph into a structured look-up table, the DT-HW compiler comprises four main steps: decision tree graph generation, tree parsing, column reduction, and ternary adaptive encoding step. We next elaborate on each of these steps.

\subsubsection{Decision Tree Graph Generation}
In this step and for a given dataset, a supervised decision tree model capable of performing multi-class classification is trained by relying on the Classification and Regression Trees (CART) algorithm \cite{breiman1984classification}. The decision tree model can be represented by a decision tree graph. The internal nodes of the graph represent rules on the attributes or features, the branches represent the decisions for the rules, and leaf nodes represent outcome classes.

\subsubsection{Tree Parsing}
Starting with a decision tree, the DT-HW compiler parses it into its equivalent table of conditions; each row in the table represents a path in the decision tree from root to leaf, and the number of rows is equal to the number of paths of the tree. Subsequently, each row consists of condition(s) applied to at least one feature.
\subsubsection{Column Reduction}
After the tree parsing step, the DT-HW compiler reduces the conditions on each feature to one single condition (or rule) per row. The incoming input features can then be easily compared against their respective features' rules. The single rule for some feature $f_i$ in row $j$, $rule_{ij}$, specifies the range for $f_i$. We note that by construct, the decision tree enforces a continuous range for the rule definition in a given path (row). The rule can be defined using a three-state comparator $\in \{$ '0' , '1' , '2', 'NaN'$\}$ and two thresholds: ($Th1_{ij}$) and ($Th2_{ij}$). The comparator states '0', '1', '2', and 'NaN' represent a-) less than or equal, b-) greater than, c-) in-between and d-) no rule for this feature in this row, respectively. In particular, if the comparator is $'0'$ in a row for some feature $f_i$, an incoming input feature, $f_{in_{i}}$, should be less than or equal to $Th1_{ij}$ (equivalently, $f_{in_{i}} \in (-Inf, Th1_{ij}]$) in order to match $f_i$'s rule in row $j$. When the comparator is $'1'$, $f_{in_{i}}$ should be greater than $Th1_{ij}$ to match the rule on $f_i$. In these two cases, $Th2_{ij}$ is ignored, and hence represented as "NaN" in the reduced table. When the comparator is $'2'$, $f_{in_{i}}$ should belong to $(Th1_{ij}, Th2_{ij}]$ in order to match the rule.
\subsubsection{Ternary Adaptive Encoding}

In the final step, the DT-HW compiler encodes each feature rule relying on an "adaptive-precision" unary encoding scheme suitable for TCAM implementations. We note that the scheme exploits the "don't care" feature of the TCAM as will be explained next. The "adaptive-precision" technique optimizes the area by setting a feature-dependent encoded string length. Thus, the number of bits varies for the different features but remains constant for a specific feature across all rows. This ensures that the encoding scheme is compact and efficient.  Hence, it is referred to as Ternary Adaptive Encoding.

The number of encoding bits for a specific feature is determined by the number of respective unique threshold values identified in the preceding column reduction step. In particular, for a given feature $f_i$ out of $N$ features ($i \in {1, 2,..., N}$), the number of bits, $n_i$, needed to encode $f_i$ depends on the number of unique thresholds over the $m$ rows, $T_i=|\cup_{j=1}^m\{Th1_{ij}, Th2_{ij}\}|$, as follows:
\begin{equation}
n_i=T_i +1
\end{equation}
Hence, for $N$ features, the total number of bits ($n_{total}$) that are eventually needed to encode the whole decision tree (excluding the leaf nodes that store the class labels) is as follows.
\begin{equation}
n_{total}=N_{branches}*\sum_i (n_i)
\end{equation}
where $N_{branches}=m$ is the number of branches or paths from the root to leaf nodes in the decision tree (or the number of leaf nodes).

The encoding scheme employs unary codes in the 'normal' form \cite{kak2016generalized}. The encoded bits belong to the basis $\{0,1,x\}$; $x$ represents a "don't care". This encoding facilitates bijective mapping of the rules into TCAM(s). The encoding can be best explained as follows for a given feature $f_i$.
\begin{enumerate}
    \item Sort the elements of  $Th^{f_i}=\cup_{j=1}^m\{Th1_{ij}, Th2_{ij}\}$ in ascending order.
    \item Construct $n_i='T_i+1'$ exclusive ranges defined in the set $R_i=\{r_1=(-Inf, min(Th^{f_i})],\, ...,\, r_n=]max(Th^{f_i}), +Inf)\}$;  
    \item Map the ranges in $R_i$ to ascending unique normal unary codes, ${u^{f_i}_{r_{1}},...,u^{f_i}_{r_{n_i}}}$, each comprising $n_i$ bits starting with the code $'00...01'$
    and ending with $11...11$.
\end{enumerate}
We note that the input features also rely on the same scheme to be encoded, and each will be represented by one of the unique feature codes based on the exclusive ranges they satisfy. 

We rely on the above encoding to construct a LUT. Recall that the rule range is continuous for a given path and thus can be interpreted in terms of the union of a set of multiple consecutive exclusive ranges. In order to accommodate for cases where a feature spans multiple exclusive ranges, we rely on "don't care" bits denoted as "x" to encode the new union range. With this scheme, inputs that belong to the different exclusive ranges that construct the rule will result in a match in the TCAM. As such, for each rule $rule_{ij}$ of $f_i$ in row $j$, we perform the following steps.
    \begin{enumerate}
        \item Find the set of exclusive ranges, $\{r_{LB}, r_{UB}\}$, spanned by $rule_{ij}$. $LB, UB \in \{1, .., n\}$. 
        \item Encode $rule_{ij}$ as follows.
        \begin{align}
        Idx=Find_{idx}(XOR(u_{r_{LB}},u_{r_{UB}})==1)
        \\
        u_{rule_{ij}}= Replace(u_{r_{LB}},Idx, "x")
        \end{align}
        $Find_{idx}(.)$ returns a list of indices satisfying a certain condition.  $Replace(u,Idx,"c")$ replaces all the characters of string $u$ in positions  $Idx$ by the character $"c"$.
    \end{enumerate}

\begin{figure}[!tb]
\centering
\includegraphics[width=0.84\linewidth,height=0.45\linewidth]
{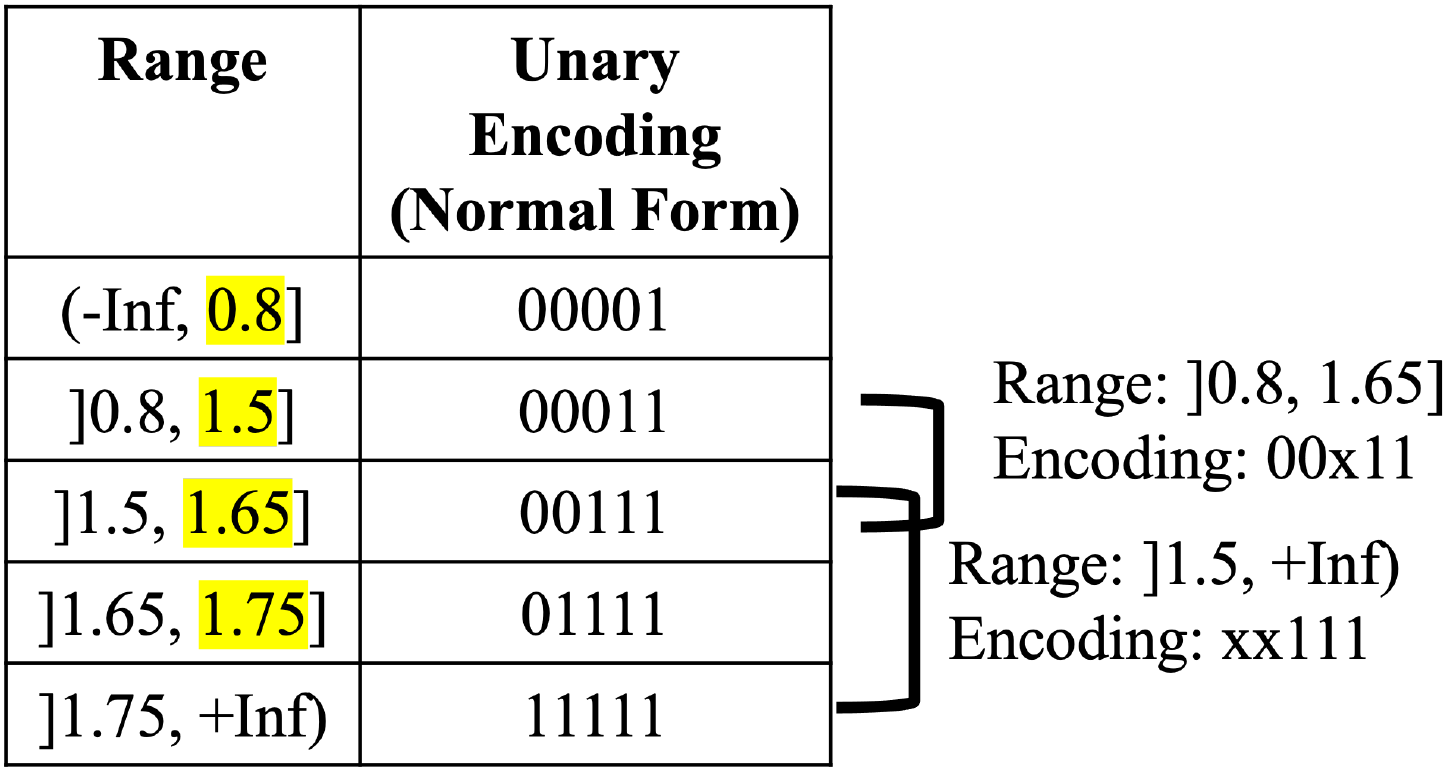}
\caption{An example of encoded ranges based on four unique thresholds (highlighted in yellow): 0.8, 1.5, 1.65, and 1.75. Unary codes in normal form are used for exclusive intervals comprising the unique thresholds. For inclusive intervals (union of multiple exclusive intervals), "don't care" bits (denotes as "x") are used to maintain the correctness of the codes of these ranges.}
\label{encodingranges}
\vspace{-0.1in}
\end{figure}

Fig. \ref{encodingranges} presents an example that illustrates the encoding scheme for some feature $f_i$. Without loss of generality, we assume that $T_i=4$ and $Th^{f_i}=\{0.8, 1.5, 1.65, 1.75\}$ as highlighted in yellow in Fig. \ref{encodingranges}. Accordingly, we construct the unary codes, \{$00001$, ..., $11111$\}, for the five exclusive ranges. We note again that we use five bits to encode each range since there are four unique thresholds. So, if in the column reduction step, $rule_{ij}="'0', 0.8, NaN"$, i.e., $f_i\leq0.8$, its range spans the first range $(-Inf,0.8]$. Accordingly, $u_{rule_{ij}}=00001$. If $rule_{ij}="'2', 1.65, 1.75"$, i.e., requiring $f_i \in ]1.65, 1.75]$, and hence it will be encoded as $u_{rule_{ij}}=01111$. To find the encoding of the new range $]0.8, 1.65]$, which spans the second and third ranges (Fig. \ref{encodingranges}), we find XOR($00011$, $00111$). This results in the string $00100$. $u_{rule_{ij}}=Replace(00011,\{'3'\}, 'x')=00x11$. In a similar manner, a range of $]1.5, +Inf[$, which spans the last three exclusive ranges in the table of Fig. \ref{encodingranges}, is encoded as $xx111$.

\begin{figure*}[!ht]
\centering
\vspace{-0.15in}
\includegraphics[width=1\linewidth]
{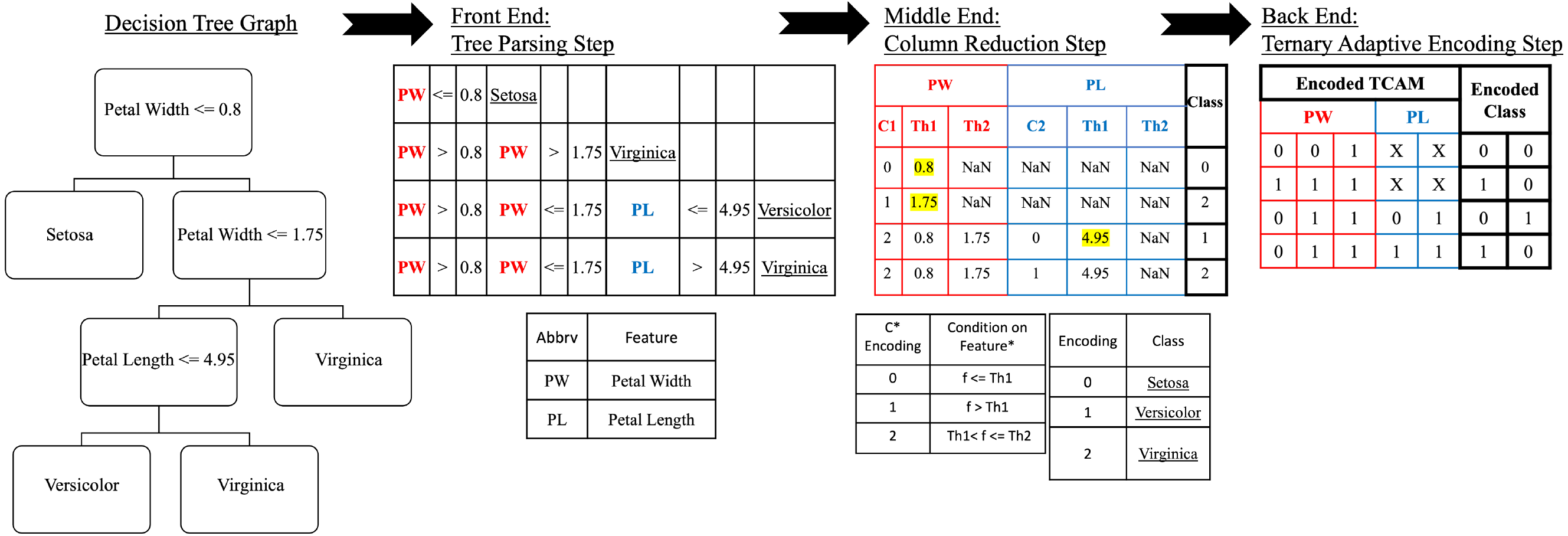}
\caption{Decision Tree Hardware (DT-HW) Compiler: Translates a decision tree graph to a structured lookup table. In particular (from left to right), it first parses the decision tree and creates a table, then reduces the columns of the table, and finally, it uses a "ternary adaptive encoding" scheme to create the look-up table.}
\label{myencoding}
\end{figure*}

\subsection{DT-HW Sample Example based on Iris Dataset} 
Henceforth, we rely on Fig. \ref{myencoding} to elaborate on the four steps described above. Starting with a given decision tree,  adapted from part of the Iris dataset  \cite{Dua:2019} decision tree, DT-HW parses it into its equivalent table of rules in the tree parsing step. The left most and right most paths in the decision tree graph are parsed into the first and second rows of the table as follows: if "Petal Width", $PW$, of the input is less than or equal to 0.8, the class output at the leaf is "Setosa" (row1). Otherwise, if the input "Petal Width" is greater than 0.8 and greater than 1.75, then the class at the leaf is "Virginica" (row2). 

Then, in the column reduction step, the second row ($PW > 0.8$ and $PW > 1.75$  Virginica) is reduced into one rule on $PW$, $PW > 1.75$ (i.e. comparator is $'1'$, $Th1=1.75$, and $Th2=NaN$). Moreover, in the column reduction step, each unique class in the original decision tree is assigned a natural number. 

By inspecting the columns of $PW$ in the column reduction step, one can notice that $PW$ has two unique thresholds so it should be encoded using three bits in the final ternary adaptive encoding step. The actual encoding follows step3 explained above. The same steps are repeated for all other rows and features. When all rules are encoded, binary encoding is further used to represent the class (decision tree leaf nodes).

\begin{figure*}[!ht]
\centering
\vspace{-0.15in}
\includegraphics[width=1\linewidth,height=0.475\linewidth]
{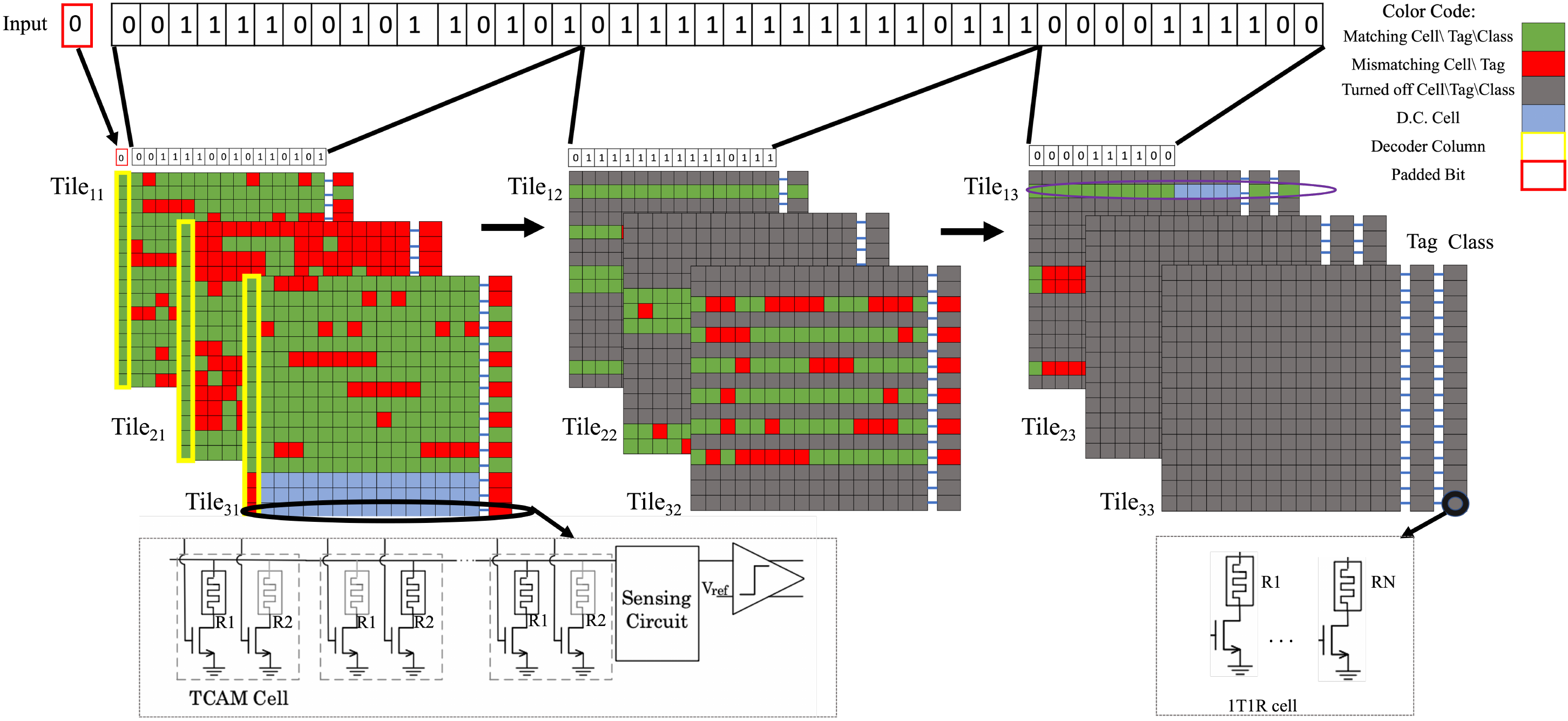}
\caption{ReCAM Functional Synthesizer: Maps the look-up table from the encoding step into ReCAM arrays and runs energy, latency, and accuracy evaluations. In other words, it breaks up, if needed, the table from the encoding step into multiple tables that can be mapped into Resistive TCAMs of regular size "$S \times S$". For purposes of energy efficiency, the column-wise TCAM tiles are separated by row-enable bits that deactivate the rows in the following tiles if the respective rows in the previous tiles mismatch.}
\label{all_framework}
\vspace{-0.1in}
\end{figure*}

\begin{figure*}[!ht]
\centering
\input{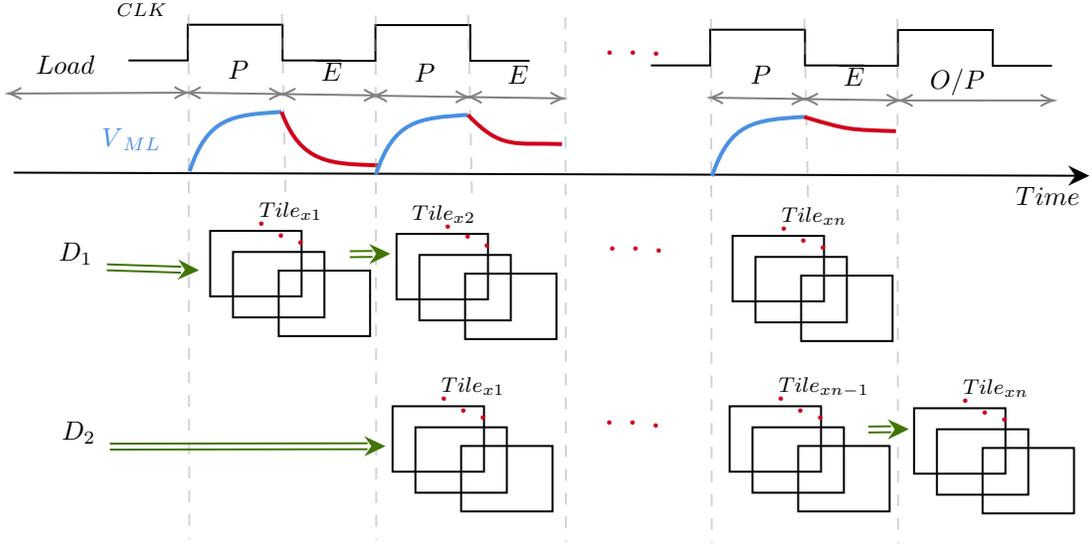}
\caption{Timing Diagram: $Tile_{x1}$ and $Tile_{x2}$ represent the row-wise tiles of the first and second column-wise TCAM tiles. P and E are short for Precharge and Evaluate respectively. $V_{in}$ is the voltage measured across $C_{in}$. First, data is loaded into the TCAM tiles. During precharge cycles, tiles are precharged, and during evaluate, the data is searched across the tiles. Column-wise TCAM tiles operate sequentially while row-wise tiles operate in parallel.}
\label{timing}
\vspace{-0.1in}
\end{figure*}

\subsection{ReCAM Functional Synthesizer}
The ReCAM functional synthesizer comprises two steps:
\begin{itemize} 
\item Mapping step: maps the look-up table, provided by the DT-HW compiler, into Ternary Resistive-CAM arrays. It takes into consideration hardware and functional limitations. 
\item Simulation step: After that, the synthesizer then evaluates energy, latency, and accuracy via simulations while maintaining certain specifications or limitations.
\end{itemize}
 
\subsubsection{Mapping}
A bit of "0", "1", or "x" in the look-up table is mapped to a "01", "10", or "11", respectively, in the two resistive elements of a TCAM cell as shown in Fig. \ref{all_framework}. Ideally, one TCAM array is used, and the total number of TCAM cells needed is equal to $n_{total}$. However, in practice, the number of TCAM cells depends on the design requirements and limitations in terms of energy efficiency, latency, and dynamic range.

\textbf{Dynamic Range:} The dynamic range of the TCAM is a limitation that needs to be satisfied to guarantee correct functionality. The dynamic range of a TCAM describes the voltage difference between a full match voltage, $V_{fm}$, and the one mismatch voltage, $V_{1mm}$, and it needs to be a "measurable difference" for the Sense Amplifier (SA) to detect it and differentiate between a full matching row scenario and a one mismatching row one. The full match voltage is that measured on the match line when all the TCAM row cells are matching, while the one mismatch voltage is the voltage measured on the match line when all cells of the TCAM are matching except one, which is mismatching. The dynamic range, $D$, is defined as follows. 
\begin{equation}
  D=V_{fm}-V_{1mm}  
\end{equation}
Furthermore, the dynamic range for a capacitive sensing design, $D_{cap}$, measured at optimal time $T_{opt}$ is defined as follows \cite{rakka2020design, bahloul2017design}.
\begin{equation}
   D_{cap}(t=T_{opt})=V_{DD}*\gamma^{(\frac{\gamma}{1-\gamma})}*(1-\gamma) 
   \label{dcap}
\end{equation}
where $V_{DD}$ is the supply voltage, and $\gamma=\frac{R_{1mm}}{R_{fm}}$, with $R_{1mm}$ and $R_{fm}$ being the equivalent resistances of the TCAM row in the cases of one mismatch and full match respectively. 

The dynamic range is affected by the TCAM row size (equivalently, the size of the row of the encoded table of Fig. \ref{myencoding}) which affects the equivalent resistances. So given a certain limit on the dynamic range, $D_{limit}$, (to render it a "measurable difference"), we calculate relying on Eqn. (\ref{dcap}) a target TCAM row size $S$ beyond which $D_{limit}$ cannot be met.

\begin{figure}[!ht]
\centering
{\includegraphics[width=0.9\columnwidth]
{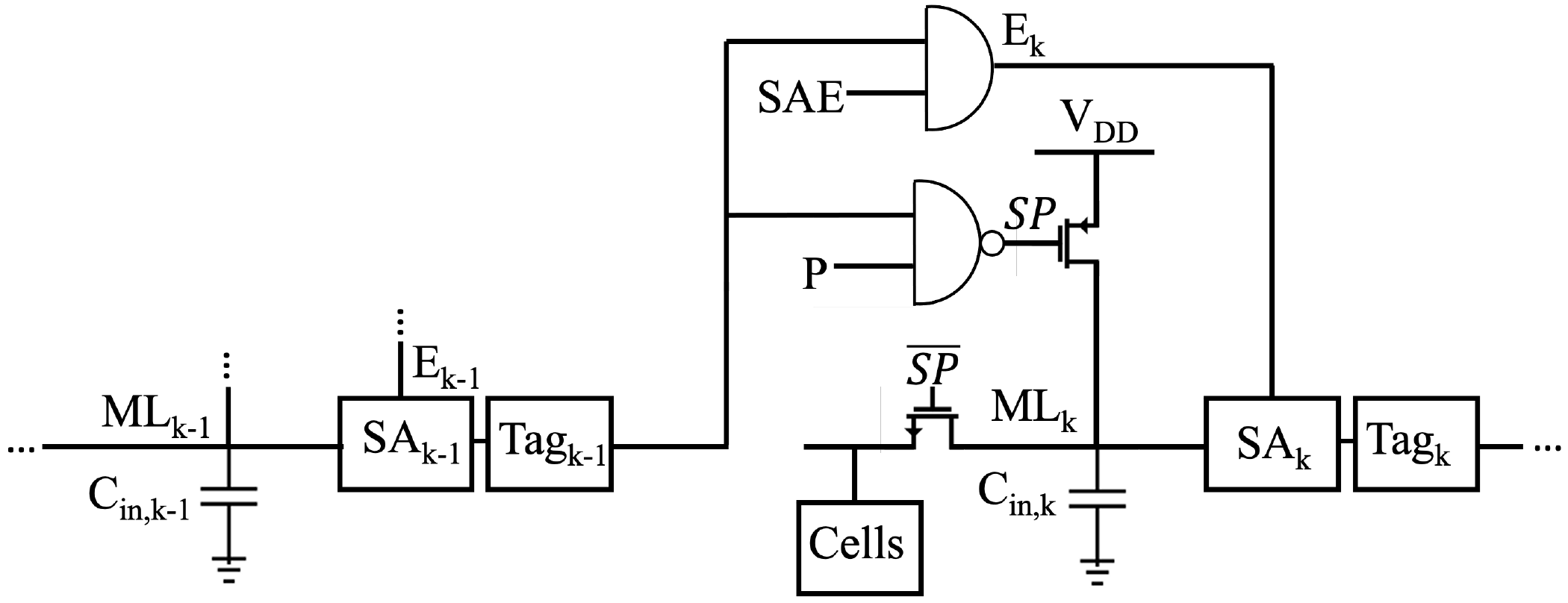}}
\caption{Selective Precharge Circuit.}
\label{SP}
\vspace{-0.1in}
\end{figure}

\textbf{Organization, Latency and Energy Efficiency:} For practical purposes, we also assume that the TCAM width (\# of rows) would be S. Hence, multiple TCAMs are needed; specifically, the synthesized TCAM cells of the encoded LUT rules need to be divided among $N_{t}=N_{cwd}
*N_{rwd}$ TCAM arrays (aka tiles) each of size $S \times S$ to guarantee practical correct operation, where $N_{cwd}=\lceil (n_{total}/\#rows\textbf{+1})/S\rceil$ and $N_{rwd}=\lceil\#rows/S\rceil$ represent the number of column-wise and row-wise TCAM tiles respectively. The $\textbf{'+1'}$ in $N_{cwd}$ is explained by the reserved decoder column discussed in the scenarios below. 
\begin{itemize}
    \item The original size of the LUT is smaller than $S \times S$ (in that case $N_{cwd}$=$N_{rwd}$=1), and the functional synthesizer needs to extend the table obtained from the encoding step by padding "don't care" cells to render the LUT size $S \times S$ (\ref{myencoding}). We reserve the first column of the TCAM array and refer to it as the decoder column to enforce mismatch for the rows that are not part of the original LUT (denoted as rogue rows).
    \item Otherwise, it needs to divide that table into multiple TCAM tiles of size $S \times S$. Tiles that are not completely filled by the LUT are padded by "don't care" cells. We reserve the first column of all TCAM arrays in the first division as decoder columns (see Fig. \ref{all_framework}.
    For purposes of energy efficiency, the column-wise TCAM tiles are separated by row-enable bits that deactivate the rows in the following tiles if the respective rows in the previous tiles mismatch. Furthermore, by setting the decoder column bits to '1' for the rogue rows, we enable further energy savings since it forcibly mismatches the rogue rows. Aside from the decoder column, the remaining columns in the rogue rows are stored as "don't care cells". 
\end{itemize}

Each one of the $S \times S$ TCAMs has a column of $S$ SAs that are used to determine the match/mismatch status of each row. The class values corresponding to the rogue rows are populated with random values from the set of possible classes. 
we equip the row-wise tiles of the last column-wise division with an extra column of ReRAM cells, that are used to store the class bits (or equivalently the encoded leaf nodes' values of the decision tree). ReRAM cells are made of 1T1R cells, and each binary bit used to encode the classes is saved in one 1T1R cell. So for a decision tree that has $C$, possible classes, $\lceil log_{2}(C) \rceil$ bits or 1T1R cells are needed for each row.  

Without loss of generality, we further elaborate on the latter scenario with the aid of Fig. \ref{all_framework}. 

\textbf{Input Processing and TCAM Mode of Operation:} A $'0'$ bit is padded at the beginning of the input. This padding along with decoder column bits enforces a mismatch in the rogue rows. For the rows that are part of the original LUT the padded bit matches with the decoder column bit. The original encoded input is then split across row-wise tiles of the column-wise tiles. Input pins that exceed the size of the encoded input may be assigned random inputs or may be masked. For the latter, the extended columns of the last column-wise division are "masked", and the "masked don't care" cells have a pair of OFF-OFF transistors and do not dissipate energy.
To exploit the parallel processing property of TCAMs whereby an input is processed in one shot across all TCAM rows, the row-wise tiles are allowed to operate in parallel. Moreover, to save precharge and evaluate energy we force a sequential operation on the column-wise TCAM tiles where no energy is dissipated in the following tiles upon mismatch in the previous tiles. The mode of operation is depicted in Fig. \ref{timing}.
Eventually, each encoded input must have one matching row in the row tiles of the last column division. We call this row the surviving row because it is the one that has matched in all corresponding previous row and column-wise tiles.

{\textbf{Selective Precharge:}}
In this work, we adopt sequential evaluation across multiple column-wise TCAM tiles for each input to enable  Selective Precharge (SP). By relying on the proposed SP circuit presented in Fig. \ref{SP}, a row that mismatches in the previous column-wise tile for a given input is not precharged nor evaluated in the current tile. In particular, if an input mismatches a given row in some $Tile_{ij}$ (stage \textit{k-1}), the SP circuit deactivates the precharge circuitry and the SA of the corresponding row in $Tile_{ij+1}$ (stage \textit{k}). Deactivating $SA_k$ prevents the floating capacitor voltage residue from falsely flagging a match and activating the following tiles while $\bar{SP}$ preserves the charge to save energy during future precharges of the same tile. As such, the advantage of using the SP circuit is depicted in the reduction of the energy-delay product presented in Fig. \ref{et-edp-c} (see Section IV for details).
We note that if an input at stage \textit{k-1} matches some row, the SA and precharge circuitry of the corresponding row in stage \textit{k} are activated.

\subsubsection{Simulation}
The synthesizer relies on simulations to carry out energy, latency, and accuracy evaluations for the design with and without hardware non-idealities. Herein, we adopt the following assumptions.

\textbf{Technology}:
For energy, latency, dynamic range ($D_{cap}$), and optimal evaluation time ($T_{opt}$) calculations, the ReCAM functional simulator relies on $16nm$ technology parameters summarized in Table \ref{parameters}.

\textbf{Target Size}:
We determine the target size $S$ values of the TCAM for $D_{limit} \in \{0.2, 0.3, 0.4, 0.5, 0.6\}$. For each $D_{limit}$ value, we rely on Eqn. (\ref{dcap}) to determine the maximum number of TCAM cells per row allowed to satisfy this value. Finally, we choose a power-of-two target $S$ value close to the maximum value found as shown in Table \ref{DR}.

\textbf{Energy:}
For energy calculation purposes, the total {energy} per an active TCAM row per an input is calculated  as follows.
\begin{equation}
    E^{active}_{row}=E_{TCAM}+E_{sa}
\end{equation}

where $E_{sa}$ is the energy of the SA obtained via SPICE simulations. In particular, for a target size $S$, $E_{sa}$ is the energy dissipated in the SA for a certain reference voltage capable of differentiating between a fully matching row and a row with one mismatch. In addition, {$E_{TCAM}$} is derived based on the closed form in \cite{rakka2020design}.
We note that the evaluation duration is $T_{opt}$, where $T_{opt}$ is the time used to sense the match line for evaluation purposes and is defined as follows.
\begin{equation}
    T_{opt}=C_{in}*ln(\frac{R_{fm}}{R_{1mm}})*\frac{(R_{fm}*R_{1mm})}{(R_{fm}-R_{1mm})}
\end{equation}
We assume the worst-case scenario for the energy calculations, where the extended cells in the row-wise tiles of the last column-wise division are treated like regular "don't care cells", hence dissipating energy as opposed to being masked. We note that we maintain the sequential functionality assuming null energy dissipation in rows that have been deactivated by the respective mismatching rows in previous tiles.

Since the energy defined above is measured per row per input, the total energy for a given input is ${E}_{total}=\sum_1^{N_a}{E^{active}_{row}}+E_{mem}$, where $N_a$ is the number of active rows for the specific input. {$E_{mem}$} is the energy needed to access the class label of the surviving row. We assume that class labels are stored in 1T1R cell(s) (total number of 1T1R cells needed is $log_{2}(\#classes)$) followed by a SA adapted from \cite{sun2018xnor}. Accordingly, $E_{mem}$ is the energy dissipated in the 1T1R cell(s) and the SA adapted from \cite{sun2018xnor}. 
The average energy per input can then be computed from all the input data points. 

\textbf{Latency:}
We define the column-wise latency, $T_{cwd}$, as  the time needed to complete the inference per input per a column-wise tile according to (\ref{latency_eqn}). The average total latency,  per input, $\bar{T}_{total}$, is then given by $\bar{T}_{total}=N_{cwd}*(T_{cwd})+T_{mem}$.

\begin{equation}
    T_{cwd}=3*(\tau_{pchg})+T_{opt}+T_{sa}
    \label{latency_eqn}
\end{equation}

where $T_{sa}$ (determined via SPICE simulations) is the time needed for the SA to sense a match or a mismatch, and {$T_{mem}$} is the time needed to access the 1T1R cell(s) storing the class label of the surviving row. We note that in the case of multiple 1T1R cells, these are accessed in parallel. In addition, our simulator operates with the maximum frequency (unless otherwise mentioned) which is given as follows.
\begin{equation}
f_{max}=\frac{1}{max((3*(\tau_{pchg})+T_{opt}+T_{sa}),T_{mem})} 
\end{equation}

This equation is used to determine the operating frequency for any array size. For instance, operating frequency for an array width of 128 is 1~ GHz under the parameters reported in Table \ref{parameters}. 



\textbf{Sense Amplifier Reference Voltage:}
We utilize two different reference voltages, $V_{ref,1}$ and $V_{ref,2}$, for the SAs. $V_{ref,1}$ is used for the SAs of all TCAM tiles except the row-wise tiles of the last column-wise division. For these tiles, $V_{ref,2}$ is utilized instead to accommodate for the presence of "masked don't care" cells that result in different values for $V_{fm}$ and $V_{1mm}$. {We note} that the sense amplifier design adopted for sensing the match line is based on the double-tail sense amplifier proposed in \cite{chiu2016double}, and is implemented in the $16nm$ technology node.

\textbf{Device Defects}
We study the accuracy-wise robustness of our DT2CAM framework under device defects. In particular, we focus on a common problem in resistive TCAM cells: the fabrication-induced permanent Stuck-At-Fault (SAF) problem. Such fault cannot be writable as it is stuck at either High-Resistance State (HRS) (equivalently stuck at the bit "0" or $SA0$) or Low-Resistance State (LRS) (equivalently stuck at the bit "1" or $SA1$) \cite{saf}. We study the SAF problem by inducing bit flips in the encoded TCAM cells as indicated in Table \ref{saf_t} and using the following probability percentage values: $SA0=[0,0.1,0.5,1,5]\%$ and $SA1=[0,0.1,0.5,1,5]\%$.

\begin{table}[b]
\centering
\caption{TCAM induced bit flips due to SAF.}
\label{saf_t}
\begin{tabular}{|c|c|c|c|}
\hline
\begin{tabular}{@{}c@{}}Target \\ Encoded Bit\end{tabular} & \{R1, R2\} & \begin{tabular}{@{}c@{}} Encoded Bit \\ w/ SA0 \end{tabular} & \begin{tabular}{@{}c@{}} Encoded Bit \\ w/ SA1 \end{tabular} \\ \hline
0 & \{HRS, LRS\} & x or 0 & 0 or \{LRS, LRS\} \\ \hline
1 & \{LRS, HRS\} & x or 1 & 1 or \{LRS, LRS\} \\ \hline
x & \{HRS, HRS\} & x & \begin{tabular}{@{}c@{}} x or 0 \\ or 1 or \{LRS, LRS\} \end{tabular}\\ \hline

\end{tabular}%
\end{table}

\textbf{Manufacturing Variability}
DT2CAM is studied (accuracy-wise) under the effect of manufacturing variability in the SAs similar to \cite{chiu2016double}. We emulate this variability by applying random offsets to $V_{ref1,2}$ of the individual SAs for a given TCAM division.  $V_{ref1,2}= \mu_{V_{ref1,2}} \pm \sigma_{sa} z *$ where $\sigma_{sa} \in [0, \, 0.03,\, 0.04,\,0.05,\,0.1]V$ and $z~N(0,1).$

\textbf{Input Encoding Noise}
To study the effect of input noise, we induce random noise in the normalized input features dataset with the following variability: $ \sigma_{in} \in [0,\,0.001,\,0.005,\,0.01,\,0.02,\,0.05,\,0.1]$ and observe the change in recognition accuracy.

{\textbf{Area}}
We estimate the average area, $A$, of the proposed synthesizer design according to the following formula.
\begin{align*}
A=N_{t}*(S^2*A_{2T2R}+S*(A_{SA}+A_{DFF}+A_{SP}))\\+S*\log_{2}(N_{c})*(A_{1T1R}+A_{SA2}) \numberthis \label{area}
\end{align*}
where $A_{2T2R}$, $A_{DFF}$, $A_{SP}$, and $A_{1T1R}$ are the areas of 2T2R (i.e. TCAM cell), D-flipflop (i.e. tag), selective precharge circuit (Fig. \ref{SP}), and 1T1R (for storing class labels for the surviving row) respectively. Moreover, $A_{SA}$ and $A_{SA2}$ represent the areas of the double-tail SA (used for sensing the match line) and the SA adopted from \cite{sun2018xnor} (used along with the 1T1R cell(s)). $N_{c}$ is the number of class labels used for some dataset. 

\section{Implementation Details}
Our framework is built in Python where we extract and parse the decision tree, then reduce it as shown in the column reduction step (Fig. \ref{myencoding}). This is followed by ReCAM functional synthesizer to perform the mapping and hardware simulations. 
To test DT2CAM, we rely on eight datasets, six of which are from the UCI Repository and Kaggle \cite{Dua:2019,kaggle1,kaggle2}. In particular, we utilize the \textit{Fisher's Iris} (denoted as Iris), \textit{Haberman's Survival} (denoted as Haberman), \textit{Car Evaluation} (denoted as Car), and \textit{Breast Cancer Wisconsin (Diagnostic)} (denoted as Cancer) datasets from the UCI repository. The \textit{Give Me Some Credit (training)} (denoted as Credit) and \textit{Pima Indian Diabetes} (denoted as Diabetes) datasets are taken from Kaggle. In addition, from Stanford's CS109 website \cite{cs109}, we utilize the \textit{Titanic} dataset. To evaluate our framework on a more recent dataset, we also test it on the \textit{COVID-19} (denoted as Covid) dataset compiled by \cite{xu2020epidemiological}. The details of the used datasets are summarized in Table \ref{tab:my-table}.

Note that in some datasets, we omit some incomplete instances and some features that are unique for each data instance (like ID or name etc.), and these modifications are reflected in the table. Without loss of generality, we use the same split percentage of the data in the aforementioned datasets to generate the decision trees: 90\% and 10\% for training and testing, respectively.

\begin{table}[!t]
\centering
\caption{Description of the utilized datasets.}
\label{tab:my-table}
\begin{tabular}{|c|c|c|c|}
\hline
Dataset & \#  Instances &\# Features & \# Classes\\ \hline
Iris & 150 & 4 & 3 \\ \hline
Diabetes & 768 & 8 & 2\\ \hline
Haberman & 306 & 3 & 2 \\ \hline
Car & 1728 & 6 & 4 \\ \hline
 Cancer & 569 & 30 & 2\\ \hline
Credit & 120269 & 10 & 2\\ \hline
Titanic & 887 & 6 & 2 \\ \hline
Covid & 33599 & 4 & 2 \\ \hline
\end{tabular}%
\end{table}

\begin{figure*}[!ht]
\centering
\subfloat[\label{et-edp-a}]{\includegraphics[width=0.65\columnwidth]
{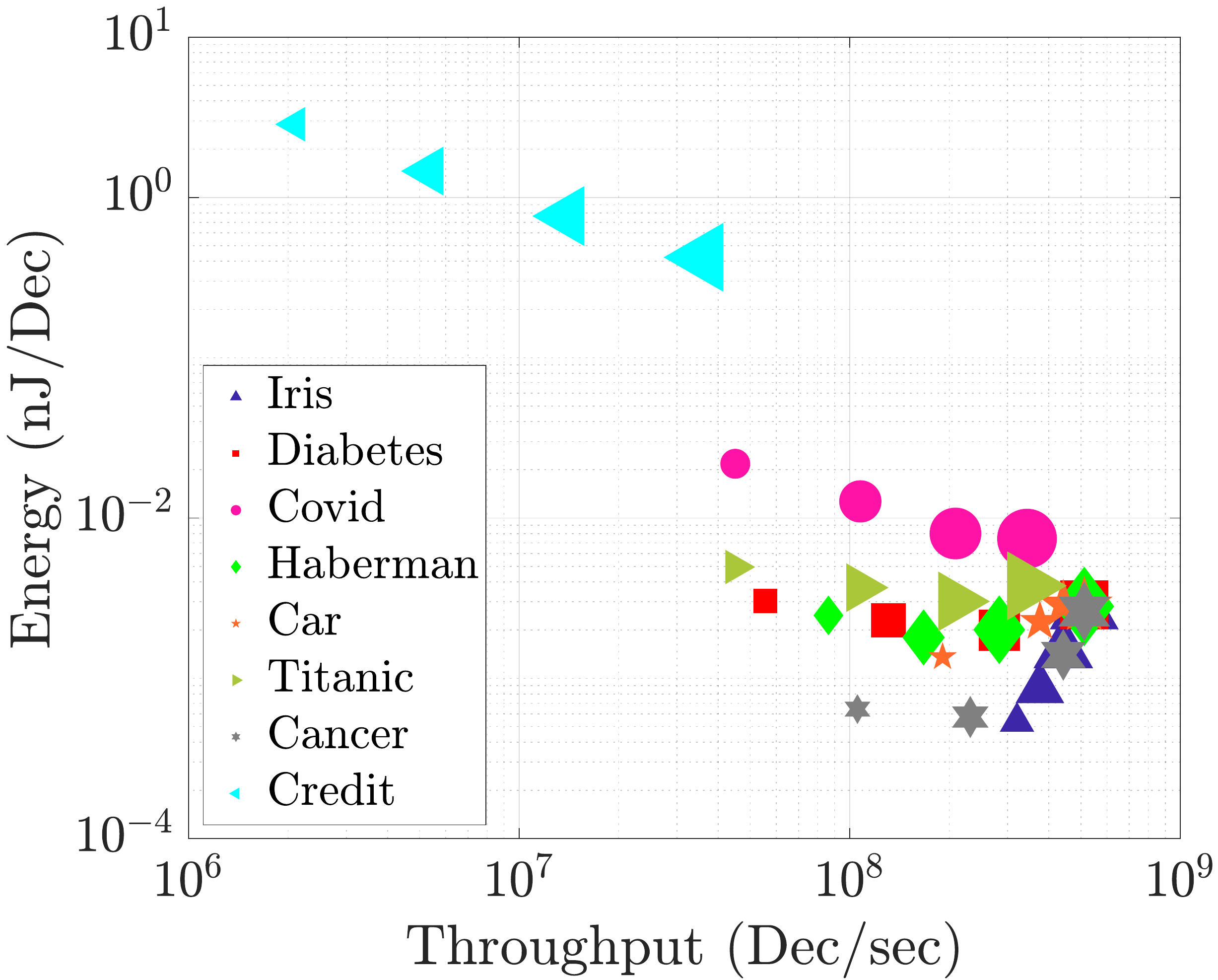}}
\hfil
\subfloat[\label{et-edp-b}]{\includegraphics[width=0.65\columnwidth]
{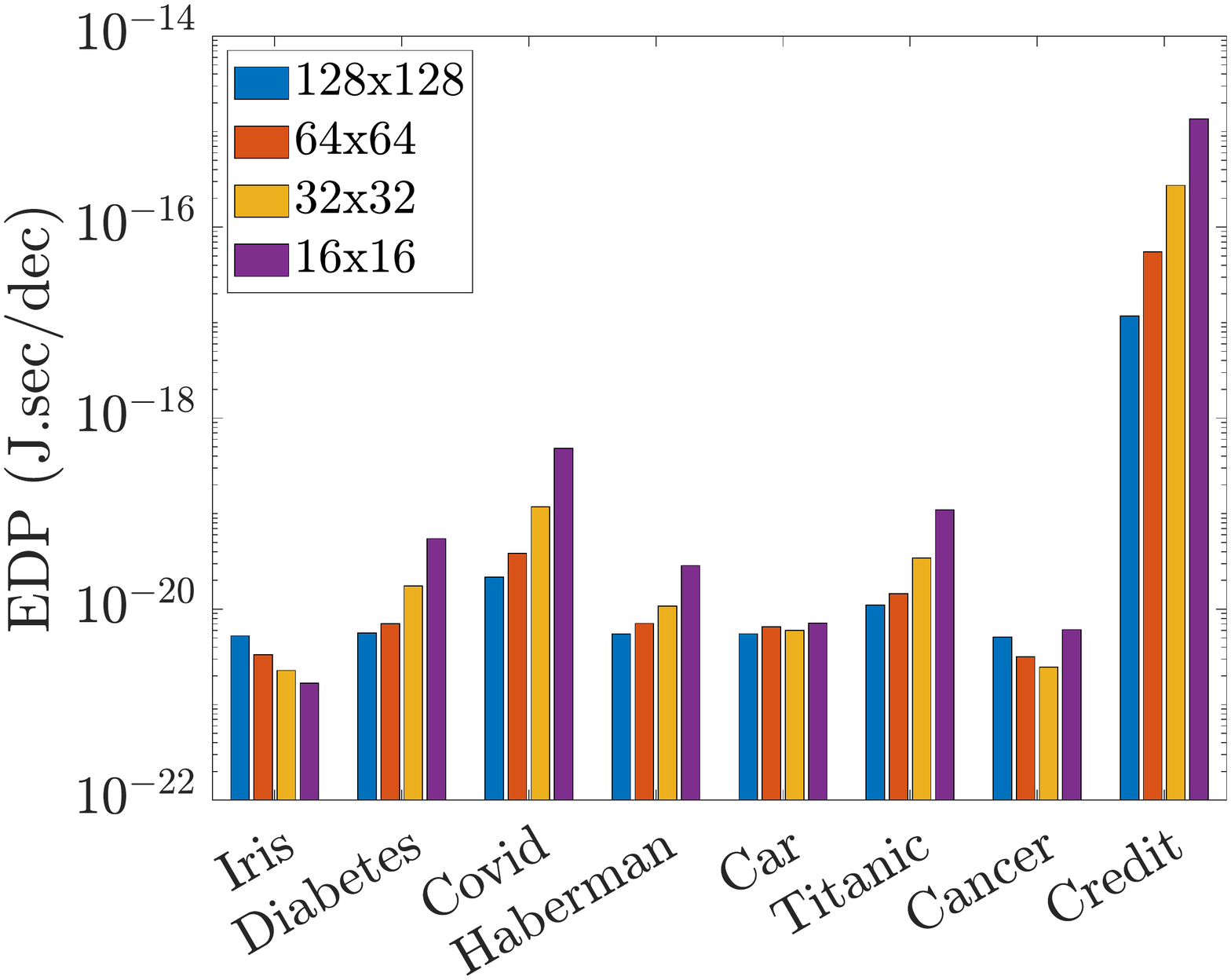}}
\hfil
\subfloat[\label{et-edp-c}]{\includegraphics[width=0.65\columnwidth]
{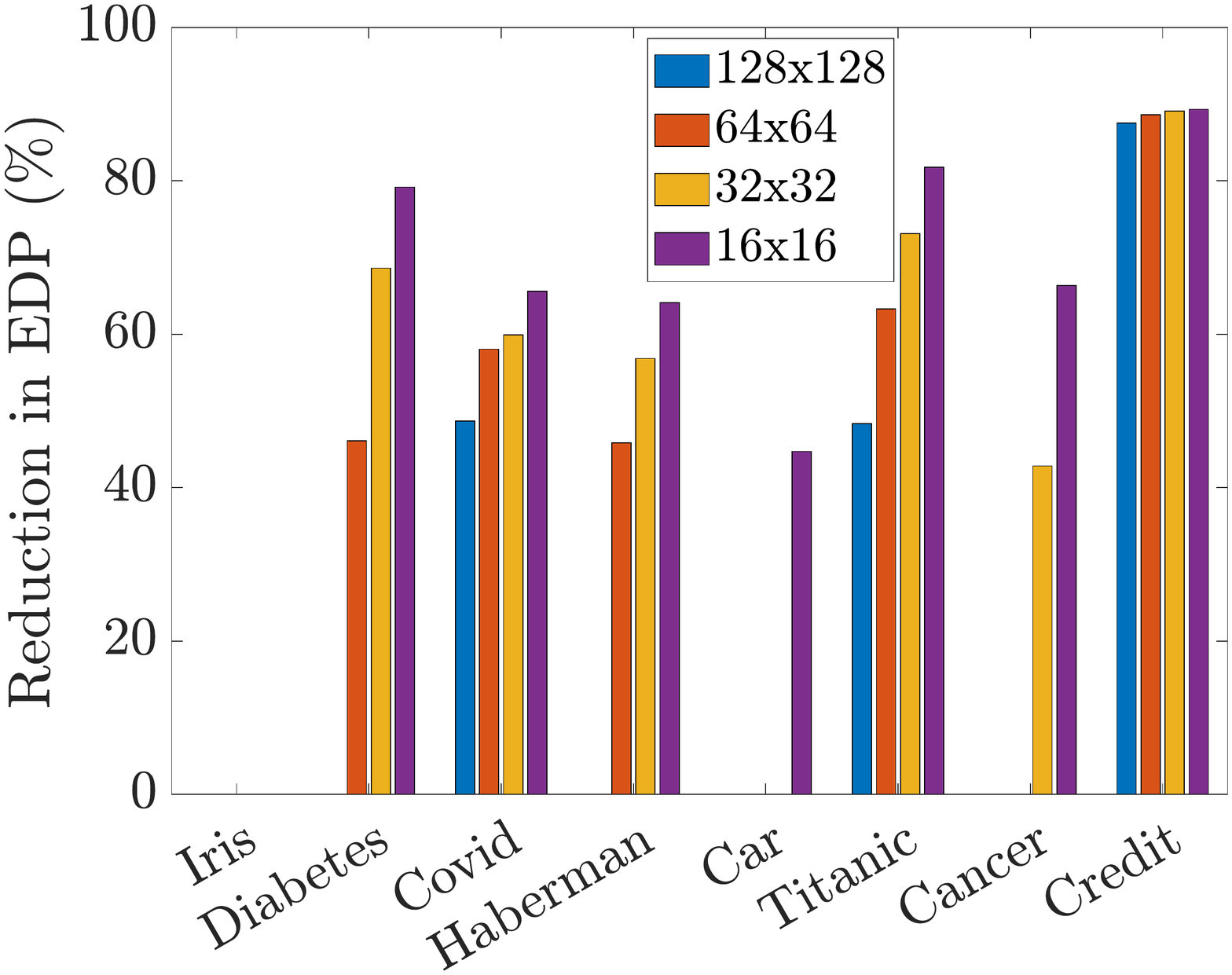}}
\caption{Per inference decision: (a) Energy vs throughput for the different datasets. The shape size determine the target size of the TCAM(s). From small to large shapes: $16\times16$, $32\times32$, $64\times64$, and $128\times128$. (b) Energy-Delay-Product, and (c) Reduction in EDP when SP is used compared to when it is not used.}
\label{et-edp}
\vspace{-0.1in}
\end{figure*}

\begin{figure*}
\centering
\subfloat[]{\includegraphics[width=0.33\linewidth]
{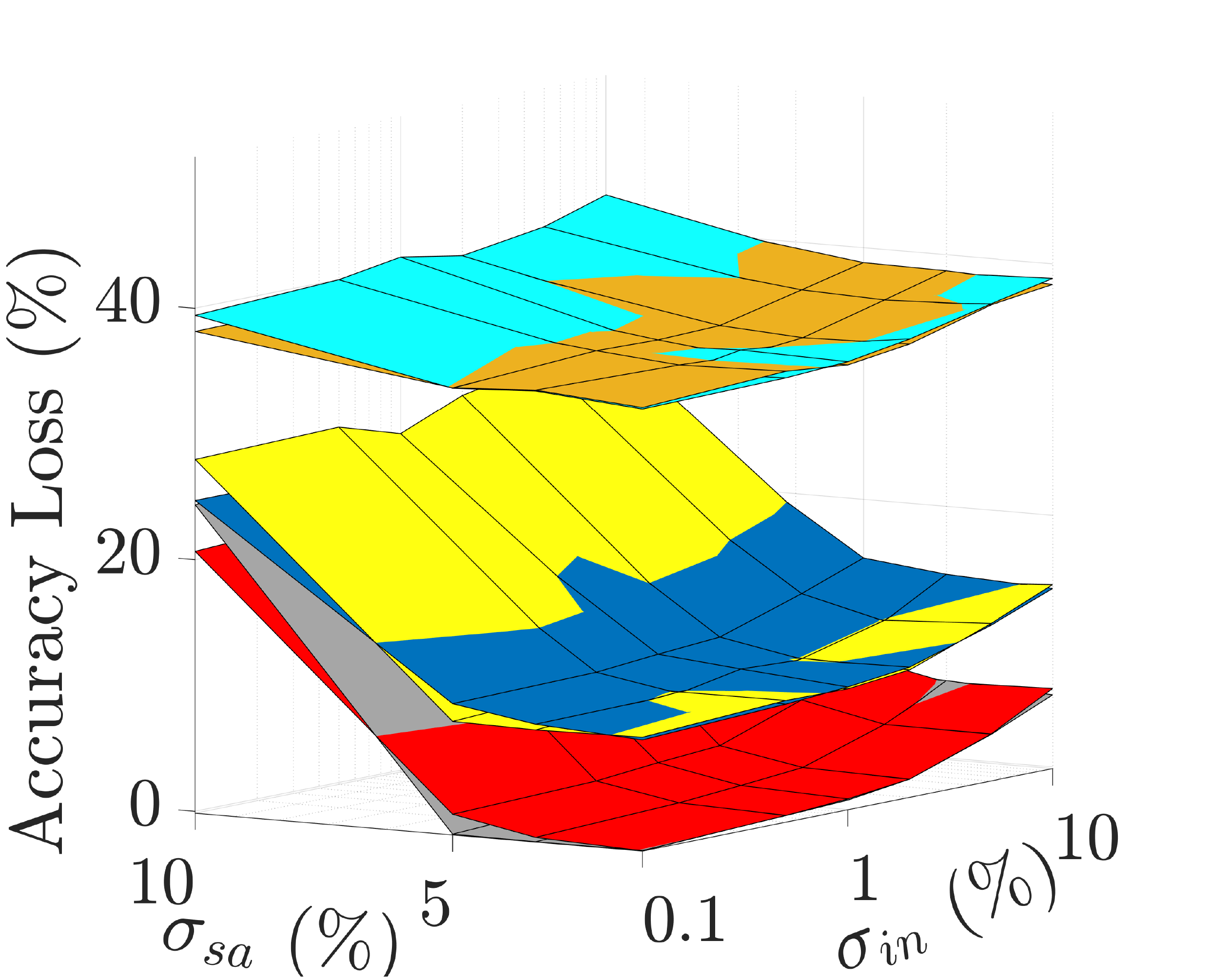}}
\hfil
\subfloat[]{\includegraphics[width=0.33\linewidth]
{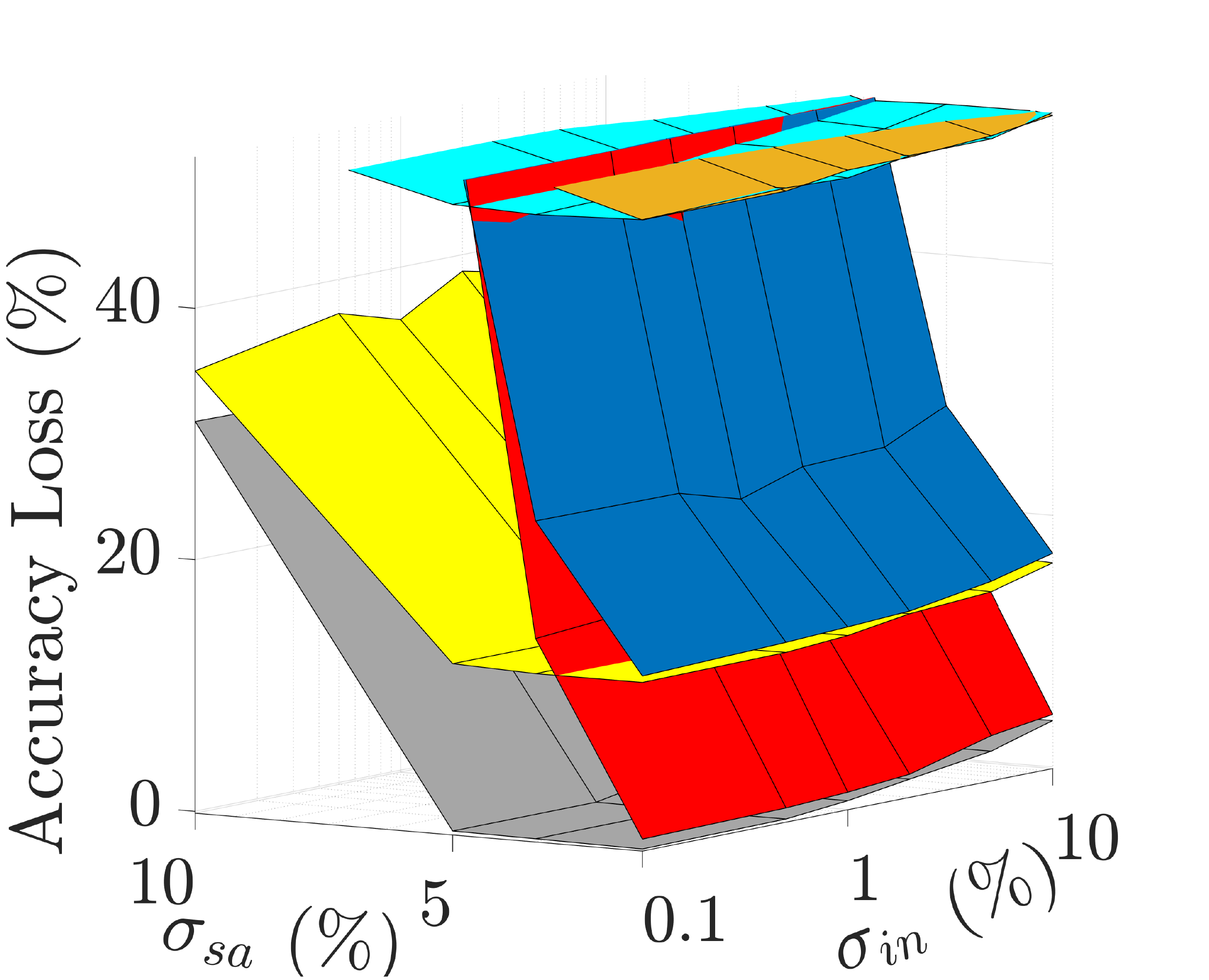}}
\hfil
\subfloat[]{\includegraphics[width=0.33\linewidth]
{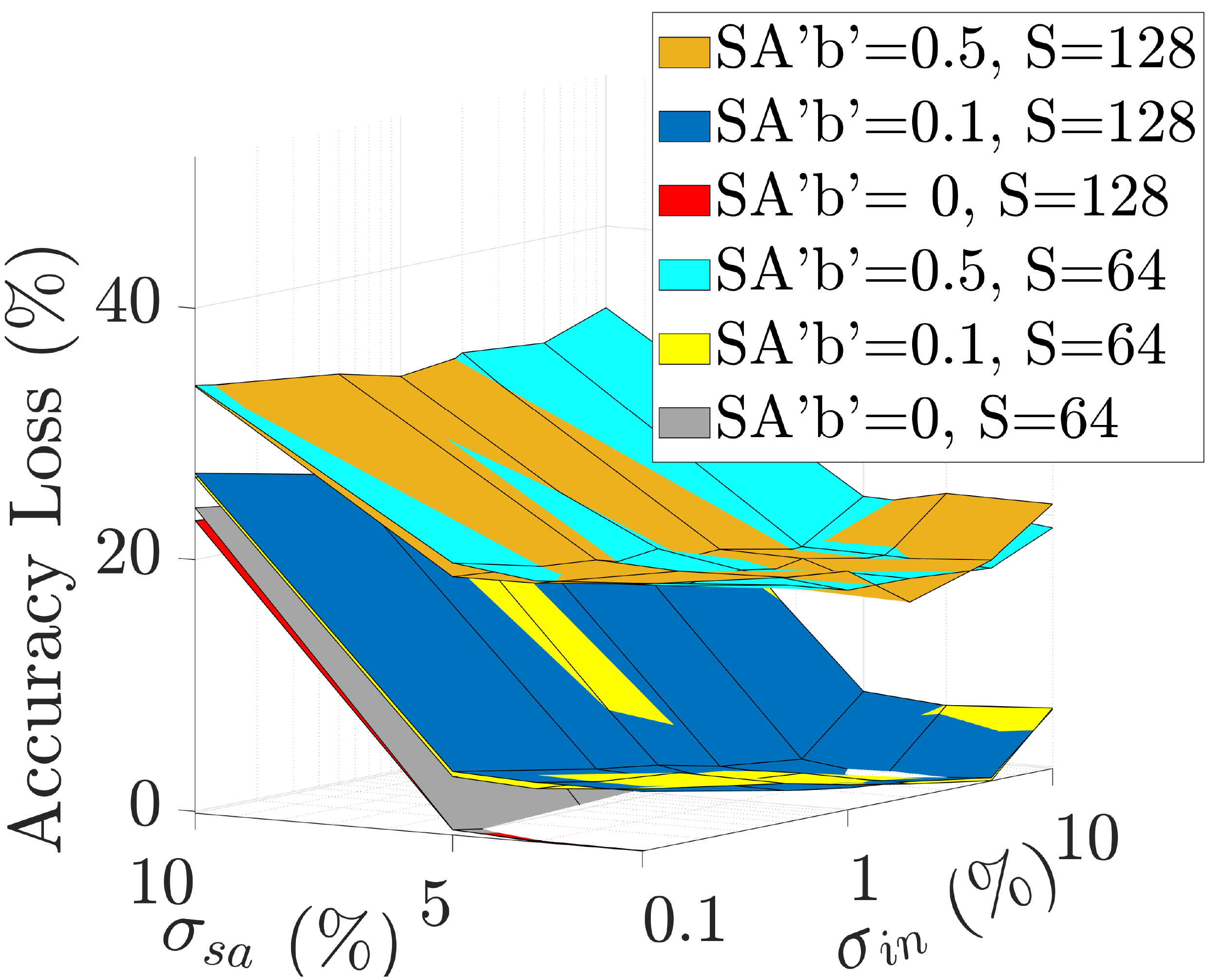}}
\caption{Percent Accuracy loss due to different hardware non-idealities (input noise, sense amplifier manufacturing variability and stuck-at-fault problem) for five datasets: (a) Diabetes, (b) Covid, and (c) Cancer. $SA'b'=x$ is equivalent to $SA0=SA1=x \%$.}
\label{results_ var}
\vspace{-0.1in}
\end{figure*}

\begin{figure}[!ht]
\centering
{\includegraphics[width=0.9\columnwidth]
{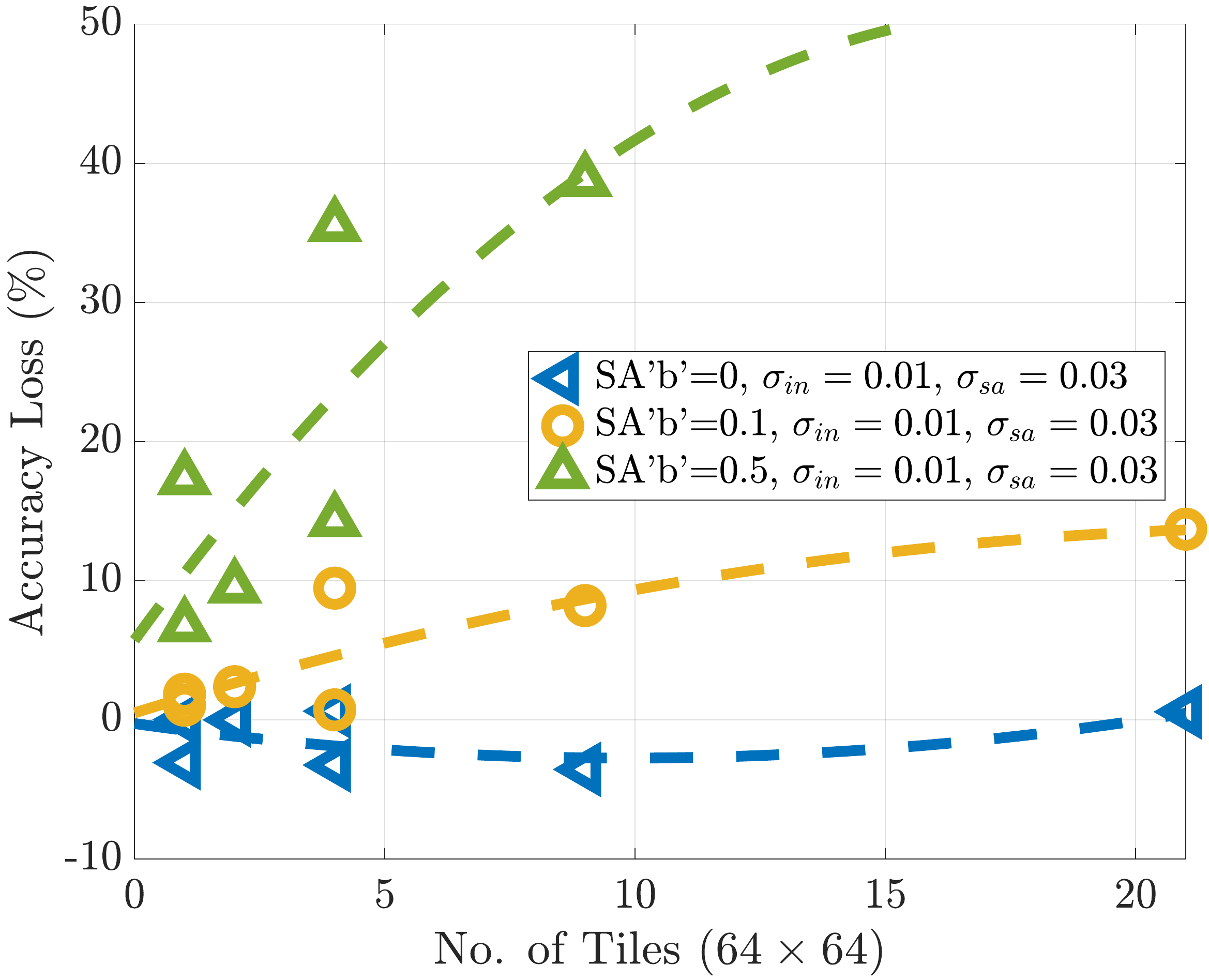}}
\caption{Accuracy Loss percentages versus the needed number of tiles. $SA'b'=x$ is equivalent to $SA0=SA1=x \%$.}
\label{accloss_tiles}
\vspace{-0.1in}
\end{figure}

\begin{table}[!t]
\centering
\caption{$16nm$ predictive technology models parameters used for the ReCAM arrays.}
\label{parameters}
\begin{tabular}{|c|c|c|}
\hline
Parameter & Definition & Value \\ \hline
$R_{LRS}$ & Low Resistance State & $5k\Omega$  \\ \hline
$R_{HRS}$ & High Resistance State & $2.5M\Omega$ \\ \hline
$R_{ON}$ & ON Transistor Resistance & $15k\Omega$  \\ \hline
$R_{OFF}$ & OFF Transistor Resistance & $24.25M\Omega$  \\ \hline
$C_{in}$ & Sensing Capacitance & $50fF$  \\ \hline
$V_{DD}$ & Supply Voltage & 1V  \\ \hline
\end{tabular}%
\end{table}

\begin{table}[b]
\centering
\caption{$D_{cap}$ values and the chosen target TCAM size $S$.}
\label{DR}
\begin{tabular}{|c|c|c|}
\hline
$D_{cap}$ Upper Bound & Max \# of Cells/Row  & Chosen $S$\\ \hline
0.2 & 154 & 128  \\ \hline
0.3 & 86 & 64 \\ \hline
0.4 & 53 & 32  \\ \hline
0.5 & 33 & 32  \\ \hline
0.6 & 21 & 16  \\ \hline
\end{tabular}%
\end{table}

\begin{table}[!t]
\centering
\caption{Number of TCAM tiles for the different datasets.}
\label{divisions}
\resizebox{0.5\textwidth}{!}{%
\begin{tabular}{cc|cccc|}
\cline{3-6}
{}                             & \multicolumn{1}{l|}{} & \multicolumn{4}{c|}{$\textbf{\# TCAM Tiles:}$ $\mathbf{N_{rwd}\times N_{cwd}}$} \\ \hline
\multicolumn{1}{|c|}{\textbf{Dataset}} & \textbf{LUT Size}     & \multicolumn{1}{c|}{$\mathbf{16\times16}$} & \multicolumn{1}{c|}{$\mathbf{32\times32}$} & \multicolumn{1}{c|}{$\mathbf{64\times64}$} & $\mathbf{128\times128}$ \\ \hline
\multicolumn{1}{|c|}{Iris}             & $9\times12$           & \multicolumn{1}{c|}{$1\times1$}            & \multicolumn{1}{c|}{$1\times1$}            & \multicolumn{1}{c|}{$1\times1$}            & $1\times1$              \\ \hline
\multicolumn{1}{|c|}{Diabetes}         & $120\times123$        & \multicolumn{1}{c|}{$8\times8$}            & \multicolumn{1}{c|}{$4\times4$}            & \multicolumn{1}{c|}{$2\times2$}            & $1\times1$              \\ \hline
\multicolumn{1}{|c|}{Haberman}         & $93\times71$          & \multicolumn{1}{c|}{$6\times5$}            & \multicolumn{1}{c|}{$3\times3$}            & \multicolumn{1}{c|}{$2\times2$}            & $1\times1$              \\ \hline
\multicolumn{1}{|c|}{Car}              & $76\times20$          & \multicolumn{1}{c|}{$5\times2$}            & \multicolumn{1}{c|}{$3\times1$}            & \multicolumn{1}{c|}{$2\times1$}            & $1\times1$              \\ \hline
\multicolumn{1}{|c|}{Cancer}           & $23\times52$          & \multicolumn{1}{c|}{$2\times4$}            & \multicolumn{1}{c|}{$1\times2$}            & \multicolumn{1}{c|}{$1\times1$}            & $1\times1$              \\ \hline
\multicolumn{1}{|c|}{Credit}           & $8475\times3580$      & \multicolumn{1}{c|}{$530\times224$}        & \multicolumn{1}{c|}{$265\times112$}        & \multicolumn{1}{c|}{$133\times56$}         & $67\times28$            \\ \hline
\multicolumn{1}{|c|}{Titanic}          & $191\times150$        & \multicolumn{1}{c|}{$12\times10$}          & \multicolumn{1}{c|}{$6\times5$}            & \multicolumn{1}{c|}{$3\times3$}            & $2\times2$              \\ \hline
\multicolumn{1}{|c|}{Covid}            & $441\times146$        & \multicolumn{1}{c|}{$28\times10$}          & \multicolumn{1}{c|}{$14\times5$}           & \multicolumn{1}{c|}{$7\times3$}            & $4\times2$              \\ \hline
\end{tabular}
}
\vspace{-0.1in}
\end{table}

\section{Results and Comparison}
In this section, we discuss the results collected by the ReCAM functional synthesizer, and then compare DT2CAM to other state-of-the-art hardware accelerators.

\subsection{Energy/Throughput/EDP Analysis}
In Fig. \ref{et-edp-a}, we plot the energy per decision (dec) vs throughput for all eight datasets and with different target $S$ value, where $ S\times S \in \{16\times16, 32\times32, 64\times64, 128\times128\}$. Larger markers indicate larger $S$ values. Inference on Credit, being the largest dataset, consumes the highest energy and has the lowest throughput, while inference on Iris, being the smallest dataset, consumes almost the lowest energy and yields the highest throughput. This is expected as energy and throughput are dataset-size dependent. For Credit, Covid, Titanic, and Diabetes (relatively large datasets as shown in Table \ref{divisions}), increasing $S$ results in reducing the per decision energy consumption (nJ/Dec) and increasing the throughput in terms of the number of decisions per second (Dec/sec). The energy reduction is due to a decrease in the number of switching blocks and SAs. The throughput improvement is attributed to the fact that the number of TCAM tiles operating sequentially for these datasets decreases with increasing $S$. Accordingly, the Energy-Delay Product (EDP) demonstrates improvement with increasing $S$ as illustrated for these datasets in Fig. \ref{et-edp-b}.

\begin{table*}[t]
\centering
\caption{Comparison with SOTA hardware accelerators. P refers to pipelined accelerators.}
\label{SOTAcomparison}
\begin{tabular}{|c|c|c|c|c|c|c|c|}
\hline
Accelerator & \begin{tabular}{@{}c@{}} Technology \\(nm) \end{tabular}& \begin{tabular}{@{}c@{}} $f_{clk}$ \\ (GHz) \end{tabular} & \begin{tabular}{@{}c@{}}Throughput \\ (Dec/s) \end{tabular} & \begin{tabular}{@{}c@{}}Energy \\ (nJ/dec) \end{tabular} & \begin{tabular}{@{}c@{}}{Area} \\ ($mm^2$) \end{tabular} &\begin{tabular}{@{}c@{}} {Area/bit}\\ ($\mu m^2$/bit) \end{tabular}
 & \begin{tabular}{@{}c@{}} {FOM}\\(J.sec.$mm^2$)\end{tabular}\\ \hline
ASIC \cite{chen2011visual} & 65 & 0.2 & $30$ & $186.7E3$ & - & - & -\\ \hline
ASIC \cite{lee2015vocabulary} & 65 & 0.25 & $60$ & $460E3$ & - & - & -\\ \hline
ASIC IMC \cite{kang201819} & 65 & 1 & $364.4E3$ & $19.4$ & - & - & -\\ \hline
ACAM \cite{pedretti2021tree} & 16 & 1 & $20.8E6$ & $0.17$ & 0.266 & 0.299 & 2.17E-18\\ \hline
P-ACAM \cite{pedretti2021tree} & 16 & 1 & $333E6$ & $0.17$ & 0.266 & 0.299 & 1.36E-19\\ \hline
DT2CAM\_128 & 16 & 1 & $58.8E6$ &  $0.098$ & 0.07 & 0.017 & 1.22E-19\\ \hline
P-DT2CAM\_128 & 16 & 1 & $333E6$ & $0.098$ & 0.07 & 0.017 & 2.15E-20\\ \hline
\end{tabular}%
\end{table*}

For the remaining datasets, the throughput (Dec/sec) improves with the target size demonstrating similar behavior as the previous ones. However, the energy consumption (nJ/Dec) increases with $S$. This is attributed to the fact that small datasets are represented by at most two tiles when $S=128$ thereby not benefiting from deactivated rows due to mismatching rows in previous tiles. Nevertheless, the throughput improvement is larger than the energy degradation (increase), and the EDP improves (decreases) with larger $S$ values (Fig. \ref{et-edp-b}). Only the Iris dataset favors smaller $S$ values when it comes to EDP due to its extremely small LUT size.

In addition, in Fig. \ref{et-edp-c}, we present the \% reduction of EDP when SP circuit is used compared to when it is not. For all datasets where at least two column-wise tiles are required for different target size $S$, we see a reduction in the EDP. This shows the advantage of using the SP circuit as it saves energy. In particular, the Credit dataset with SP circuit achieves the highest reduction in EDP (around 90\%). This is expected as it is the largest dataset with the largest produced LUT, which in turn yields a large number of column-wise tiles. The large number of column-wise tiles benefits from the SP circuit by evaluating only few rows in each tile.  

\subsection{Analysis with Hardware Non-idealities}
We study the DT2CAM framework in the context of accuracy loss for different target size $S$, and under the described hardware non-idealities in section II-C. Without loss of generality, we focus on the following datasets: Diabetes, Cancer, and Covid. We note that for all the datasets under study, the accuracy evaluated by the ReCAM synthesizer for ideal hardware (without non-idealities) matches the \textit{accuracy obtained in Python} (hereon denoted as \textit{golden accuracy}) when inference is performed. Hence the accuracy loss of each dataset is measured concerning the corresponding golden accuracy. From Fig. \ref{results_ var}, the target size $S$ does not impact the accuracy loss in the presence of non-idealities for Diabetes and Cancer. For the Covid dataset, which has a large number of tiles, a smaller $S$ is more robust against non-idealities as the drop in accuracy is lower. This is clear for the case when $SA'b'=0.1\%$ and $S=64$ (yellow plane) and $S=128$ (dark blue plane). The same holds for the case of $SA'b'=0\%$. Note that the cases for $SA'b'=0.5\%$ are truncated for better illustration. Note that the probability of a defect falling in a division decreases with $S$. The variability induced in SAs affects the accuracy more severely compared to the noise in the input test datasets, and this applies to all datasets under study. In fact, for some cases, the input noise reduces the accuracy loss, and this is due to the test dataset itself, and how it changes with the induced input noise. We finally note that the stuck-at-fault problem affects the accuracy the most, as it can increase the \% accuracy loss up to 50\% (in the absence of other non-idealities), especially for large $S$.

\begin{figure}[!t]
\centering
{\includegraphics[width=0.9\columnwidth]
{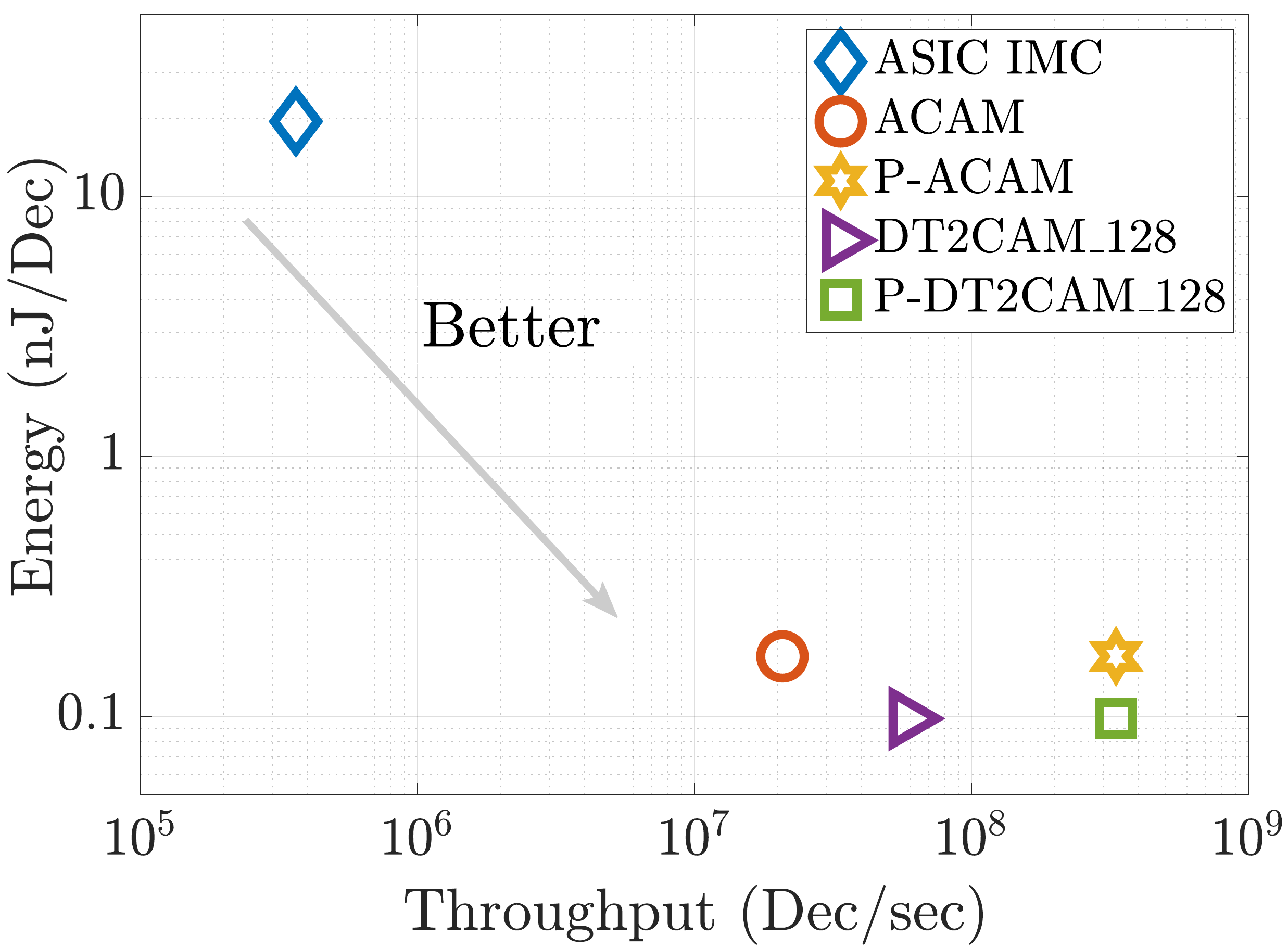}}
\caption{Energy vs. Throughput for our proposed DT2CAM and other SOTA hardware accelerators.
}
\label{energy_throughput_sota}
\vspace{-0.1in}
\end{figure}

\subsection{{Comparison with Other Hardware Accelerators}}
In Table \ref{SOTAcomparison}, we summarize the per decision throughput and energy for our framework and other hardware accelerators for decision tree inference (\cite{ chen2011visual,pedretti2021tree, kang201819,lee2015vocabulary}). For DT2CAM, we assume a 2000x2048 original TCAM size, divided into 128x128 ($S=128$) tiles to mimic inference on the traffic dataset problem. In particular, we take into consideration the 2000 rows by 256 features reported for the traffic dataset in \cite{pedretti2021tree}, and further assume that each feature will require eight bits of storage (overestimation). We report the values for the sequential case (column-wise tiles operate sequentially) and pipelined case (column-wise tiles are pipelined). Compared to the ASIC accelerators (\cite{chen2011visual,lee2015vocabulary}) and IMC based accelerators (ASIC IMC \cite{kang201819}, ACAM \cite{pedretti2021tree}), 
our proposed DT2CAM achieves the highest throughput ($58.8E6$  $Dec/s$) and consumes the lowest energy ($0.17 nJ/Dec$). Furthermore, our proposed pipelined design has the same throughput while consuming $1.73x$ lower energy than the pipelined ACAM design \cite{pedretti2021tree}.

Furthermore, we report the average area of DT2CAM (based on Eqn. (\ref{area})), and we report the average area per bit (i.e. $A/\#TCAM Cells$) in Table \ref{SOTAcomparison}. Compared to the area reported for the analog CAM framework \cite{pedretti2021tree}, we achieve about $3.8x$ and $17.5x$ reduction in area overhead and area/bit respectively.

We define a figure of merit, FOM, to better compare the accelerators' performances as follows. 
\begin{equation}
    FOM= EDP*A
\end{equation}
Accordingly, the lower the FOM (i.e., smaller energy-delay product and area ), the better the performance. Our sequential/parallel DT2CAM framework has 17.8x /6.3x  better FOM compared to the ACAM realization. 


\section{{Conclusion}}
In conclusion, we proposed DT2CAM, a decision tree to the ReCAM framework which is capable of evaluating the energy, latency, and accuracy of performing decision tree inference using TCAMs (resistive in particular) with and without hardware non-idealities. The proposed framework comprises two main phases: the DT-HW compiler which maps a decision tree graph into a look-up table and the ReCAM functional synthesizer which maps the look-up table into ReCAM arrays and performs simulations. Experiments on various datasets with varying the number of features and complexity show that the ternary adaptive encoding scheme adopted by the DT-HW compiler is robust against noise and efficient in terms of energy and latency. Compared to other SOTA hardware accelerators, DT2CAM achieves the lowest energy, highest throughput, lowest area overhead, and lowest FOM (preferred).  It is also worth mentioning that our framework, including selective precharge, can be extended to accommodate other ReRAM cell typologies, including ACAM \cite{bazzi2022efficient}, resulting in better performance which we consider for our future work.  



\end{document}